\documentclass[nofootinbib,aps,pre,twocolumn,superscriptaddress]{revtex4-1}

\usepackage{graphicx}
\usepackage{epstopdf}
\usepackage{amsmath}
\usepackage[usenames,dvipsnames]{color}
\usepackage{adjustbox}
\usepackage{tabularx}

\usepackage[normalem]{ulem}

\begin{document}

\title{Rejuvenation and Shear-Banding in model amorphous solids}

\author{Armand Barbot}
\author{Matthias Lerbinger}
\affiliation{PMMH, CNRS, ESPCI Paris, Universit\'e PSL, Sorbonne Universit\'e,
 Universit\'e de Paris, 75005 Paris, France}
\author{Ana\"{e}l Lema\^{i}tre}
\affiliation{Universit\'{e} Paris-Est, Laboratoire Navier (UMR 8205), CNRS, ENPC, IFSTTAR, F-77420 Marne-la-Vall\'{e}e, France}
\author{Damien Vandembroucq}
\author{Sylvain Patinet}
\email[]{sylvain.patinet@espci.fr}
\affiliation{PMMH, CNRS, ESPCI Paris, Universit\'e PSL, Sorbonne Universit\'e,
 Universit\'e de Paris, 75005 Paris, France}

\date{\today}

\begin{abstract}
We measure the local yield stress, at the scale of small atomic regions, in a deeply quenched two-dimensional glass model undergoing shear banding in response to athermal quasistatic (AQS) deformation. We find that the occurrence of essentially a single plastic event
suffices to bring the local yield stress distribution to a well-defined value for all strain orientations, thus essentially erasing the memory of the initial structure. It follows that
in a well-relaxed sample, plastic events cause the abrupt (nucleation-like)
emergence of a local softness contrast and thus precipitate the formation of a band,
which, in its early stages, is measurably softer than the steady-state flow.
Moreover, this postevent yield stress ensemble presents a mean value
comparable to that of the inherent states of a supercooled liquid around the
mode-coupling temperature $T_{\rm MCT}$. This, we argue, explains that the transition between brittle and ductile yielding in amorphous materials occurs around a comparable parent temperature. Our data also permit to capture quantitatively the
contributions of pressure and density changes and demonstrate
unambiguously that they are negligible compared with the changes of
softness caused by structural rejuvenation.
\end{abstract}


\maketitle

\section{Introduction}

Shear banding, an intense localization of plastic strain within narrow bands, is the primary mode of mechanical failure in a variety of amorphous materials including metallic glasses~\cite{greer_shear_2013} or granular media~\cite{le_bouil_emergence_2014}. When loading conditions permit, bands may extend throughout a piece of material and sustain repeated plastic activity while preserving their overall structure~\cite{pampillo_comprhensive_1974}. While it is thus clear that the material inside shear bands is softer than the surrounding, undeformed, glass, the structural origin of this softness constrast, and the mechanisms underlying its rapid emergence, remain highly debated topics~\cite{lewandowski_temperature_2006,Schuh-ActaMat07,martens_spontaneous_2012,ketov_localized_2013,nicolas_effects_2016}.

Progress on these issues is hampered by basic difficulties in describing the mechanisms of plastic deformation in glasses~\cite{spaepen_microscopic_1977,sollich_rheology_1997,falk_dynamics_1998,manning_localization_2007,shi_evaluation_2007,Greer-NatRevMater16,tong_mechanical_2018}.
In these systems, indeed, structural disorder disallows the existence of topological defects akin to crystalline dislocations~\cite{mecking_kinetics_1981,ding_soft_2014}; and plasticity results from the intermittent occurrence of local rearrangements (``flips'') triggered when the atoms within ``zones'' (small regions a few atoms wide) reach mechanical instabilities~\cite{maloney_universal_2004,Tanguy_EPJ06,lemaitre_plastic_2007}.
Although the proximity of a packing to instabilities may correlate with usual observables such as local density, pressure, or shear stress, these correlations are weak~\cite{tsamados_local_2009,patinet_connecting_2016}, and the question remains open of their predictive value in the context of shear banding.

Mesoscopic models~\cite{tyukodi_depinning_2016,nicolas_deformation_2018} of plasticity represent a glass as an elastic continuum in which zones are embedded. They prove able to reproduce most of the phenomenology of amorphous plastic deformation, in particular the appearance of shear bands, yet only by postulating that local plastic thresholds are weakened after the occurrence of local rearrangements~\cite{vandembroucq_mechanical_2011,tuzes_disorder_2017,castellanos_avalanche_2018,popovic_elasto-plastic_2018}. No atomistic data, however, exist that support this idea.

In this article, we use a recently developed numerical method~\cite{puosi_probing_2015,patinet_connecting_2016,barbot_local_2018} that permits to access local yield stress at the zone scale (i.e. in regions a few atoms wide) in arbitrary orientations. This method was previously applied to quenched glasses, prior to any plastic deformation. In a companion paper~\cite{patinet_bauschinger_2019}, we use it to show that the Bauschinger effect  results from plasticity-induced yield strength anisotropies. Here, we use this technique to resolve the local yield stress during shear banding. We thus demonstrate that the occurrence of a small amount of plastic strain erases the memory of the local structural properties acquired during aging. Moreover, it brings the average local yield stress, in all shear orientations, to a value smaller than in the steady flow state. The mean yield stress of postevent zones compares with that of supercooled liquid inherent states (ISs) around $T_{\rm MCT}$: this explains that the equilibration temperature separating ductile and brittle responses is located in the vicinity of $T_{\rm MCT}$~\cite{ozawa_random_2018}. When loading a well-relaxed sample, the suddenness of the rejuvenation process causes a rapid drop in the local yield stress which precipitates the formation of a band, that persists over large strain scales thanks to the softness contrast.

\section{Numerical methods}

We use the same atomistic model as in Ref.~\cite{barbot_local_2018}: a two-dimensional binary system comprising $10^4$ equal-mass atoms that interact via a Lennard-Jones (LJ) potential with second-order smoothing at cutoff. All values are given in LJ units. Preparation and deformation are performed with periodic boundary conditions at a constant volume corresponding to a density $\rho\simeq 1.02$. Samples are prepared via a slow temperature ramp at rate $\dot T=0.32\cdot10^{-6}$ across the glass transition temperature range. This allows for equilibration until the alpha relaxation time is of order $T_g/\dot T\simeq10^{6}$, which corresponds to a fictive temperature $T_{f}\simeq 0.29$ as detailed in Appendix~\ref{appendix_temperature}. When reaching $T=0.078~T_{\rm MCT}$, each sample is further quenched by energy minimization to obtain a mechanically equilibrated state. These initial states are then subjected to athermal quasistatic (AQS)~\cite{maloney_universal_2004,maloney_amorphous_2006} simple shear deformation along the $x$ axis up to a linear strain $\gamma=5$, with steps $\Delta\gamma=10^{-4}$ and a minimization force tolerance of $10^{-11}$. Unless otherwise specified, all averages and distributions are obtained by collecting data from $100$ independent samples.

In such strongly deformed systems, local strains must be computed using finite-strain theory (see Appendix~\ref{appendix_strain}). A scalar measure of local strain $\epsilon$ is provided by the square root of the second invariant of the logarithmic (Hencky) strain. The macroscopic counterpart of this quantity is $E=\ln(\gamma/2+\sqrt{1+\gamma^2/4})$.

\begin{figure}
\includegraphics[width=0.95\columnwidth]{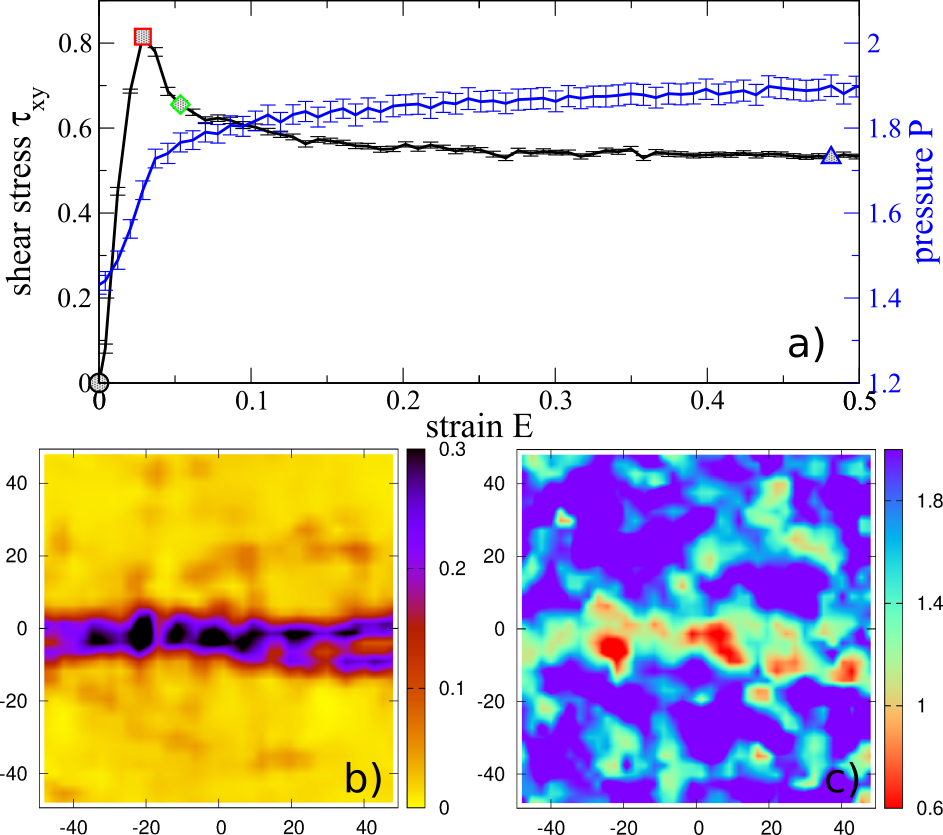}
\caption{\label{fig:stress_strain} a) Mean shear stress $\tau_{xy}$ (black) and pressure $P$ (blue) as a function of Hencky's equivalent shear strain $E$ during simple shear AQS loading. Bottom: local maps at $E=0.05$; b) accumulated strain $\epsilon$, and c) yield stress $\tau^{c}_{xy}$.}
\end{figure}

The sample-averaged stress-strain response during constant-volume AQS loading is reported in Fig.~\ref{fig:stress_strain}a. It displays a typical stress overshoot followed by softening. As observed in other systems~\cite{utz_atomistic_2000}, macroscopic pressure $P$ rises sharply until the peak and continues to grow, albeit more slowly, thereafter. This pressure increase in our constant-volume simulations is the analog of a volume increase in typical fixed-pressure experiments. It is the signature that plastic activity creates configurations that are less well packed than the original aged ones, an idea usually invoked to justify the still classical free-volume theory. We will, however, rule out this interpretation quantitatively by showing that the plastically-induced pressure increase, or local density decrease, contributes only a minor fraction of strain softening.

The plastic response of our relaxed systems systematically involves shear banding as illustrated in Fig.~\ref{fig:stress_strain}b: Local strain field maps, as shown here at $E=0.05$, typically display a high-strain region crossing the cell throughout. This phenomenon may only arise from structural causes, since our use of AQS shearing rules out the competition between timescales~\cite{martens_spontaneous_2012}, inertia~\cite{nicolas_effects_2016} or local heating~\cite{lewandowski_temperature_2006}.

Local yield stress values are measured in sheared configurations using
the method of
Refs.~\cite{patinet_connecting_2016,barbot_local_2018}. It consists in
isolating circular inclusions of radius $R_{\rm free}=5$ and
straining them by requiring outer atoms to follow an affine motion
corresponding to pure shear along an arbitrary orientation
$\alpha\in[0,\pi]$. Inner, inclusion atoms are free to move nonaffinely and
required, using the AQS method, to remain mechanically balanced. The
first plastic rearrangement they undergo determines the yield point of
the inclusion in orientation $\alpha$ and defines the local yield stress
$\tau^{c}(\alpha)$ and critical strain $\epsilon_{c}(\alpha)$. For any
considered $\alpha$, the yield threshold is thus identified for all
inclusions centered at regular grid
points. In this article, we primarily focus on the changes in softness that affect the forward barriers, in the loading orientation with $\alpha=0$, due to their direct relevance to the forward plastic response. Yet, we will show [Section \ref{subsec_iso}] that softening occurs in all shear orientations, and in particular in the backward one ($\alpha=\pi/2$).

Let us emphasize that we focus here on the threshold $\tau^{c}$, not on the residual strength $\Delta\tau^{c}=\tau^{c}-\tau$, with $\tau$ the local stress. $\Delta\tau^{c}$ is an adequate predictor of the immediate plastic response~\cite{patinet_connecting_2016,barbot_local_2018}, but explicitly depends on local stress, i.e. on elastic loading. $\tau^{c}$ does not and thus constitutes an intrinsic gauge of local glassy structures.

\section{Local yield stress evolution}

\subsection{Shear band profiles}

Figure~\ref{fig:stress_strain}c presents the resulting local yield
stress map in the loading orientation
$\tau^{c}(\alpha=0)=\tau^{c}_{xy}$, at $E=0.05$.
When comparing with the strain map of Fig.~\ref{fig:stress_strain}b, it is
strikingly clear that the material in the band tends to present lower
yield stress values than the material around it. We thus directly see here that, as expected, plasticity induces softening and thus localization in the band~\cite{rice_localization_1976}. Below we use this ability to gain insight on the mechanisms underlying the band formation and stability.

\begin{figure}
\includegraphics[width=0.95\columnwidth]{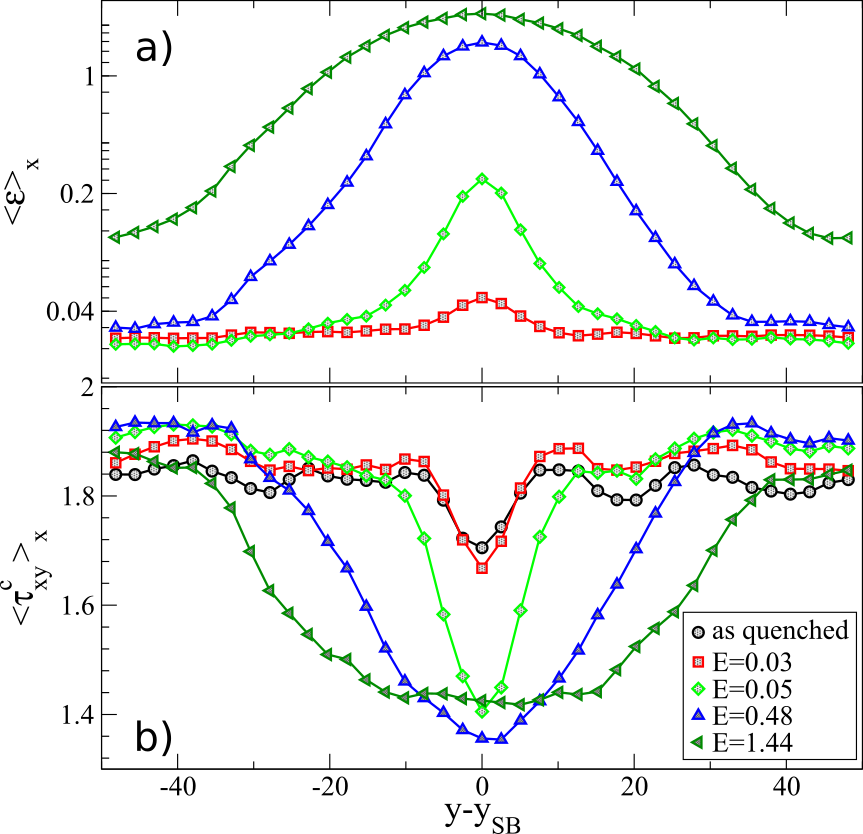}
\caption{\label{fig:shear_band} Local averages computed along the shear direction $x$ for different strain levels $E$ as a function of the vertical position $y$ with respect to the shear band position $y_{SB}$: a) strain $\langle\epsilon\rangle_{x}$ and b) yield stress $\langle\tau^{c}_{xy}\rangle_{x}$. The symbols used for the different strain levels are the same as for Fig.~\ref{fig:stress_strain}a.}
\end{figure}

To quantify the emergence and evolution of bands, we focus first on samples ($53$ out of $100$) where a single band nucleates horizontally~\footnote{In $25$ samples, a band initially nucleates vertically, yet cannot persist due to the simple shear Lees-Edwards boundary conditions: A horizontal band then emerges later. The rest of the samples exhibit more complex behavior and show intermediate crossed configurations, porelike structure~\citep{alix-williams_shear_2018} and multiple shear bands.}. We identify as band center the ordinate $y_{SB}$ where the $x$-averaged local strain reaches its maximum, at the macroscopic strain $E=0.1$ where the band is clearly visible in all samples. We then report in Fig.~\ref{fig:shear_band}, as a function of $y-y_{SB}$, the $x$-averaged strains and yield stresses at several macroscopic strains.

The strain profiles [Fig.~\ref{fig:shear_band}a] clearly show that the band grows over time~\cite{shi_evaluation_2007,alix-williams_shear_2018}. This would be inaccessible in experiments as the band evolution is usually interrupted by mechanical failure~\cite{greer_shear_2013}. But, in simulations, since periodic boundary conditions maintain the system integrity, the softness contrast eventually resorbs itself and the system reaches a unique, statistically homogeneous, steady flow state~\cite{patinet_bauschinger_2019}. The yield stress profiles (Fig.~\ref{fig:shear_band}b) distinctly show that the softer region coincides with the band and grows at a similar rate. At $E=0.05$, the contrast of local yield stresses between the band and the outer region is of order $27\%$, in striking agreement with experimental hardness measurements in metallic glasses~\cite{pan_softening_2011}.

Similarly band-centered density, shear modulus, and potential energy profiles are displayed in Appendix~\ref{appendix_localprop}. They show that the band also presents a $2\%$ lower density, a $17\%$ lower shear modulus and a $5\%$ higher potential energy than the surrounding material. In agreement with Ref.~\cite{hinkle_coarse_2017}, these profiles demonstrate that the position of the band correlates, albeit at different degrees, with the initial values of the considered properties in the quenched state. Although the initial central fluctuation would decrease with system size, this merely illustrates that the AQS response is completely governed by structure. Yet, the question remains to understand why softening occurs just after a few percentage strains, thus promoting the formation of the band.

\subsection{Local yield stress vs local plastic strain}

\begin{figure}[t]
\includegraphics[width=0.95\columnwidth]{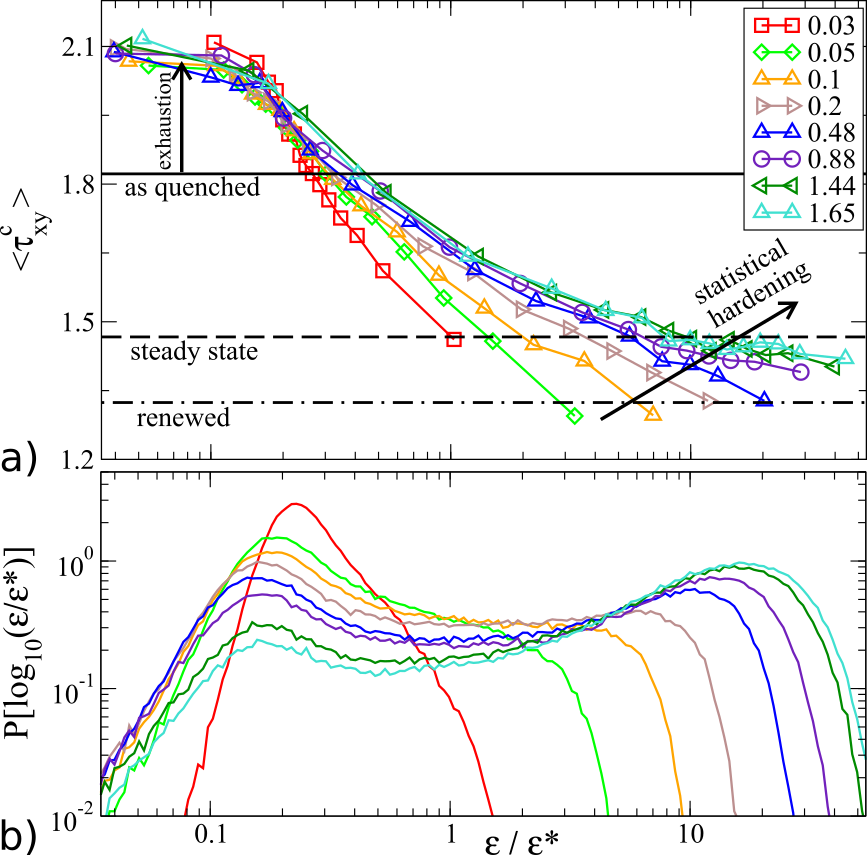}
\caption{\label{fig:mean} Top: Average yield stresses $\langle\tau^{c}_{xy}\rangle$ as a function of the normalized local strain $\epsilon/\epsilon^{*}$ for various macroscopic strain $E$. The continuous, dashed and dash-dotted horizontal lines correspond to $\langle\tau^{c}_{xy}\rangle$ in the as-quenched, steady state and renewed ensembles, respectively. Bottom: distribution of $\log_{10}\left(\epsilon/\epsilon^*\right)$.}
\end{figure}

In this perspective, we report in Fig.~\ref{fig:mean}, for different $E$'s, the average yield stress (top) conditioned by the local strain $\epsilon$ from the quench state. These data are collected from all the $100$ systems of our ensemble (irrespective of the initial band orientation). The distribution of $\log_{10}\left(\epsilon/\epsilon^{*}\right)$, which is the abscissa of this lin-log graph, is reported in the bottom panel and shows a characteristic two-peak structure~\cite{lemaitre_plastic_2007} that separates low-strain, elastically responding regions with the high strain ones, which have undergone local plastic events. As $E$ increases, the first peak expectedly decreases in amplitude, while the second one grows and shifts rightward, which is expected to result from the accumulation of plastic activity. The crossover strain scale between the two peaks happens to be $\epsilon^{*}\simeq2\langle\epsilon_{c}\rangle$, with $\langle\epsilon_{c}\rangle=0.054$ the average critical strain in the as-quenched state. $\epsilon^{*}$ thus corresponds roughly with the scale of the local strain change caused by a plastic rearrangement.

The $\langle\tau^{c}_{xy}\rangle$ curves of Fig.~\ref{fig:mean}a constitute the first \emph{direct and quantitative} observation of local softening associated with local yielding. They systematically decay with $\epsilon$, with a characteristic strain of order $\epsilon^{*}$. This key observation entails that the memory of the initial state is erased with a characteristic strain corresponding to essentially one event. It explains the rapidity of softening and thus localization. Yet, there are several features of these curves that are surprising and call for further explanations: (i) the very low strain ($\epsilon/\epsilon^{*}\lesssim0.1$), elastic regions present higher yield stress values than the quenched state; (ii) there is a significant spread between the curves, which entails that the local yield stress depends on both $E$ and $\epsilon$; and more specifically, (iii) at small $E$, highly strained regions are unexpectedly softer than the steady (homogeneous) flow state, so that at a given $\epsilon$, the local yield stress grows with $E$, which constitutes a hardening effect.

\begin{figure}
\includegraphics[width=0.95\columnwidth]{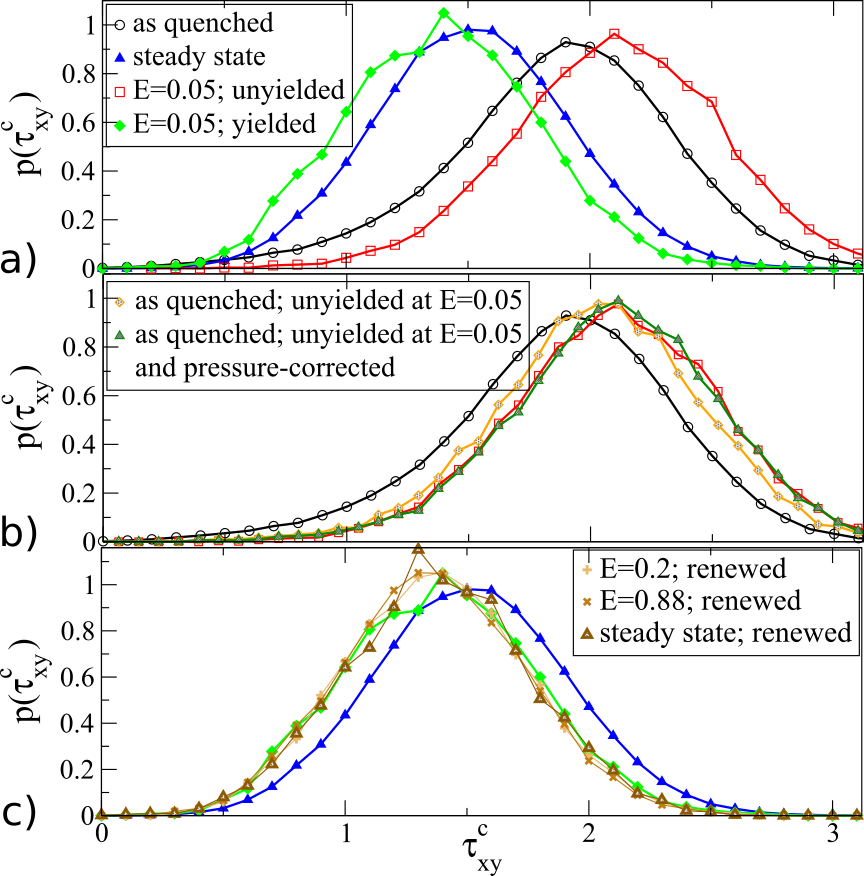}
\caption{\label{fig:pdf} a) Distribution of local yield stresses $\tau^{c}_{xy}$ in the as-quenched state (black), the steady flow state (blue), at $E=0.05$ in the unyielded (red) and yielded (green) areas and the renewed state (shades of brown). b,c) examine the effect of pressure and statistical hardening (see text for details), for the as-quenched state and unyielded sites (b); the steady flow state and yielded sites (c).}
\end{figure}

To clarify these issues, we focus on $E=0.05$, and distinguish within each configuration two types of local environments: (a) regions where $\epsilon<\epsilon^{*}/8$, are called ``unyielded''; and (b) those where $\epsilon>2\epsilon^{*}$, are called ``yielded''. The distributions of local yield stresses in these two subsets are shown in Fig.~\ref{fig:pdf}, along with their counterparts in the initial as-quenched ensemble and steady homogeneous flow. We find that: (A) unyielded regions (mainly outside the shear band) are harder than the as-quenched state and (B) yielded ones are (unexpectedly) softer than the steady flow. The hardening and softening previously noted above under points~(i) and~(iii) (respectively)~\footnote{{These effects can also be seen in the $\langle\tau^{c}_{xy}\rangle_{x}$ profiles of Fig.~2b, as material outside the band initially hardens, while the middle of the band softens at largest $E$'s.}}, hence, affect the whole yield stress distribution; in fact, they even occur in all orientations as shown below in Fig.~\ref{fig:pdf_iso}.

The hardening of unyielded sites can be attributed primarily to a statistical effect: the progressive elimination (exhaustion) from this ensemble of the weakest sites of the quenched state when they yield~\cite{baret_extremal_2002}. Indeed, if we pick out the unyielded sites at $E=0.05$, and plot their yield stress distribution in the initial quenched state [Fig.~\ref{fig:pdf}b, orange diamonds], then we recover the major part of the hardening. The residual difference is entirely attributable to the large (nearly $0.4$) pressure increase during loading. Indeed, in unyielded sites, $\langle\tau^{c}_{xy}\rangle$ depends roughly linearly, but quite weakly, on $P$ (see Appendix~\ref{appendix_localprop}); and when both exhaustion and the linear pressure dependence of $\langle\tau^{c}_{xy}\rangle$ are taken into account (green triangles), we recover the distribution of yield stress outside shear bands.

We previously observed that yielded sites at finite strains (especially $E=0.05$) are weaker than the steady flow state.
We now show this results from a general property of ``freshly renewed'', i.e., postyield sites. To evidence this idea, we pick out sites that yield (i.e., present a local strain change $\Delta\epsilon>2\epsilon^{*}$) over a fixed, small amount of macroscopic strain $\Delta E=0.05$ (as before), yet starting from different initial states, say strain $E$.
Figure~\ref{fig:pdf}c compares the yield stress distributions in these post-yield states for $E$ values ranging from $0$ (as-quenched) to the steady AQS flow state. Strikingly, all these distribution collapse.
This constitutes a key observation of our work. It establishes that essentially one plastic event brings the local structure to a unique yield stress distribution, regardless of the initial state.

Note that, the final states explored in Fig.~\ref{fig:pdf}c present sufficiently different pressure levels that we should see systematic deviations from collapse~\footnote{Our renewed ensembles show mean pressure differences up to $0.25$ which are expected to correspond to mean yield stress differences $\approx0.0625$}. This is not the case, which implies that the pressure-yield-stress relation is specific to each renewed ensemble as shown in Appendix~\ref{appendix_localprop}: The renewal process does not produce a unique structural ``state'', but instead a unique yield stress distribution under different conditions. This idea is often speculated to hold in the construction of
mean-field~\cite{bouchaud_weak_1992,sollich_rheology_1997,hebraud_mode_coupling_1998} or meso-scale
models~\cite{vandembroucq_mechanical_2011}, but was never directly observed. It entails that shear banding results
from a process akin to nucleation: the production of structures of a specific softness level after essentially one plastic event.

The hardening of yielded sites with increasing $E$ then  appears to occur due to the rapid elimination, via further yielding, of the weakest among the freshly renewed sites. As $E$ increases, beyond $0.05$, the regions of a fixed $\epsilon$, comprise an increasing fraction of sites that have yielded early on, but are strong and have resisted since. This finally explains that the steady flow state is harder than the freshly renewed sites.

\subsection{Local yield stresses in other directions \label{subsec_iso}}

\begin{figure}[h]
\includegraphics[width=0.95\columnwidth]{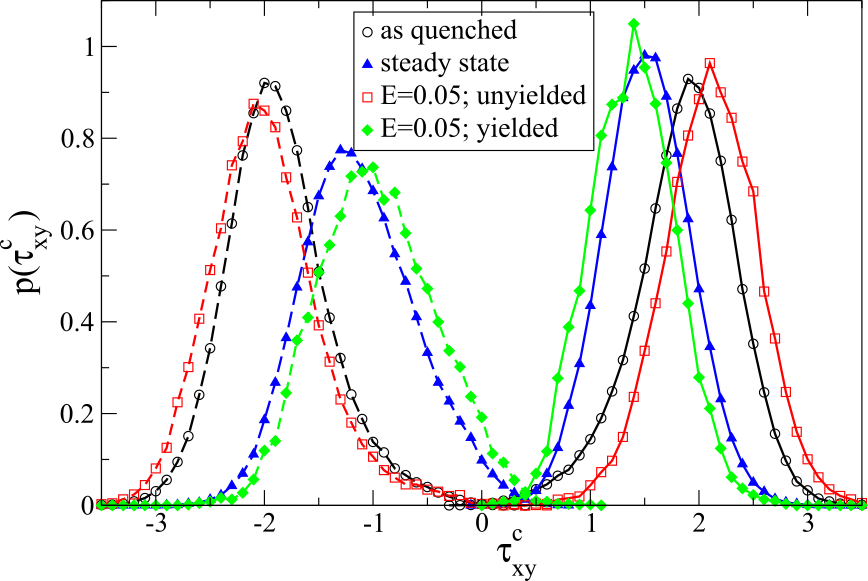}
\caption{\label{fig:pdf_iso} Distribution of local yield stresses $\tau^{c}_{xy}$ in the loading (continuous lines) and reverse directions (dashed lines) in the as-quenched state (black), the steady flow state (blue), at $E=0.05$ in the unyielded (red) and yielded (green) areas.}
\end{figure}

The distributions of local yield stresses in the forward ($\alpha=0$) and backward in ($\alpha=\pi/2$) global shear directions are shown in Fig.~\ref{fig:pdf_iso}. As for the forward simple shear direction investigated above, we report the yielded and unyielded states at $E=0.05$ , along with their counterparts in the initial as-quenched ensemble and steady homogeneous flow. Here, $\tau_{xy}^c=\tau^c$ for $2\alpha=0$ and $=-\tau^c$ for $2\alpha=\pi$, so that forward (respectively reverse) barriers corresponds to positive (respectively negative) abscissa. The lack of $\tau_{xy}^c\to-\tau_{xy}^c$ symmetry in all three deformed ensembles points to a flow-induced anisotropy studied in detail in Ref.~\cite{patinet_bauschinger_2019}. We do not address this effect in this article, but focus on the overall changes in softness that affect the forward barriers in the loading orientation. We see here that the rejuvenation process, in particular the strain-softening, occurs in fact in all orientations~\cite{patinet_bauschinger_2019}.

\subsection{Local yield stress vs free volume}

\begin{figure}[h]
\includegraphics[width=0.95\columnwidth]{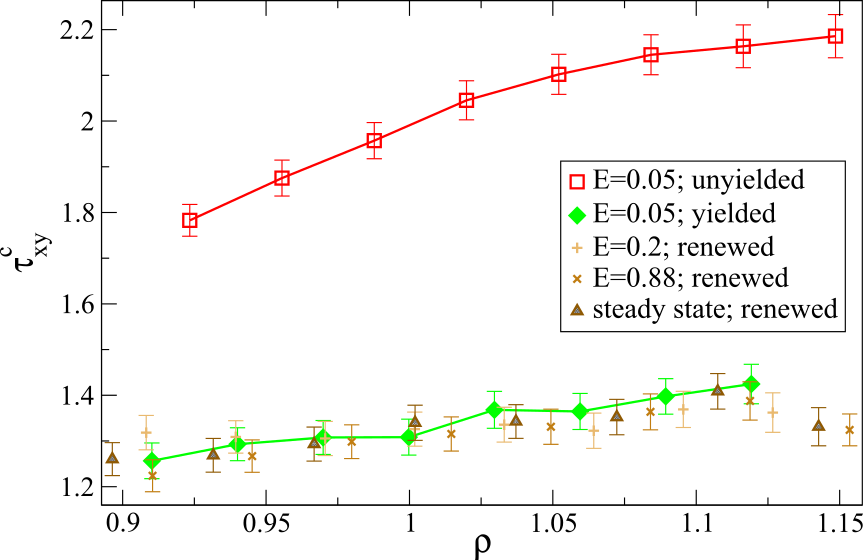}
\caption{\label{fig:mean_vs_density} Average local yield stress $\langle\tau^{c}_{xy}\rangle$ as conditioned by local density in unyielded (red) and yielded (green) states at $E=0.05$ and in the renewed ensembles (shades of brown).}
\end{figure}

To quantify the possible role of local expansion in softening, we report in Fig.~\ref{fig:mean_vs_density} $\langle\tau^{c}_{xy}\rangle$ as conditioned by local density in both yielded and unyielded site. These data show very clearly that, at the same $\rho$, yielded and unyielded sites present sharply different yield stresses: local density, hence, cannot predict local softness. The $\simeq2\%$ density difference seen between the band and the outer, elastic, region corresponds to a maximum $\tau^{c}_{xy}$ variation of $\simeq4\%$, which is insignificant compared with the observed $\simeq27\%$ change in $\tau^{c}_{xy}$. The leading cause of softening is the production by plasticity of packings in a different ``state'', i.e., presenting different $\tau^{c}_{xy}$ vs $\rho$ (or other) relation, than the initial material.

\subsection{Variation of local yield stresses with parent temperature}

\begin{figure}[h]
\includegraphics[width=0.95\columnwidth]{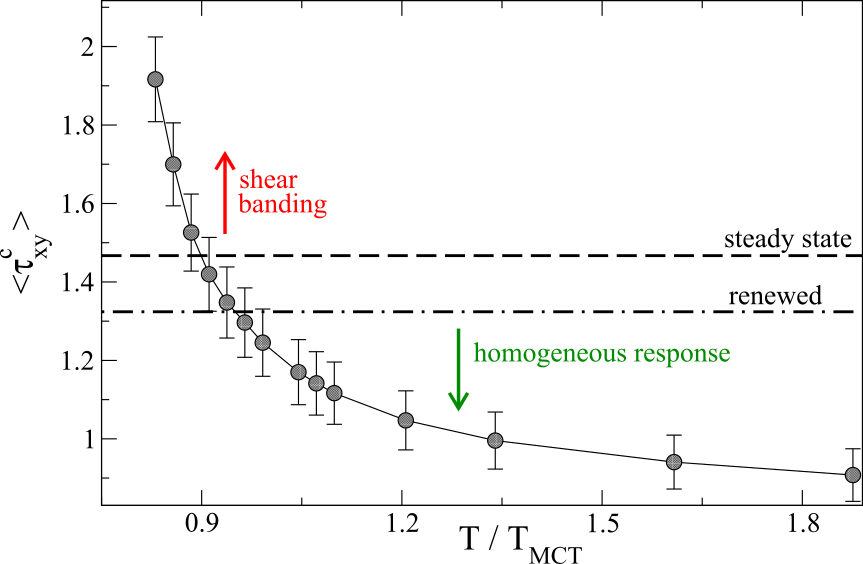}
\caption{\label{fig:mean_vs_T} Average local yield stresses $\langle\tau^{c}_{xy}\rangle$ of inherent states as a function of their parent (liquid) temperatures.}
\end{figure}

We report in Fig.~\ref{fig:mean_vs_T} the average $\langle\tau^{c}_{xy}\rangle$ computed for inherent states obtained after instantaneous quenches from different parent temperatures of equilibrated liquids. It is noteworthy that the value $\langle\tau^{c}_{xy}\rangle\simeq 1.32$ in postyield states is comparable with the average yield stress in the ISs of a supercooled liquid at $0.9~T_{\rm MCT}$, a range of temperatures where the dynamics of liquids enters the activated regime. This indicates that postyield structures have comparable barriers heights with a liquid lying on the upper layers of the potential energy landscape (PEL). From this standpoint, the fact that the average yield stress in postyield sites compares with that of liquid ISs near $T_{\rm MCT}$ appears to constitute a fundamental feature of the PEL. It then explains that the critical parent temperature separating brittle and ductile yielding transitions in amorphous materials is located in the vicinity of $T_{\rm MCT}$~\cite{ozawa_random_2018}, since a contrast must exist between the yield stress in the initial state and in the postyield sites for shear banding to occur.

Note, however, that we consider here just an average. Furthermore, rearranged and inherent states cannot be strictly compared due to the nonzero polarization in the barrier ensemble along the loading direction~\cite{patinet_bauschinger_2019} and to the slight increase of the pressure in flowing states. Despite this oversimplification, an inherent state obtained near $T_{\rm MCT}$ is therefore expected to give a good estimate of the stationary threshold distributions, and thus act as a limit between softening (leading to strain localization) and hardening regimes in deeply and poorly quenched glasses, respectively~\cite{ozawa_random_2018}. This result is also fully consistent with recent findings obtained in oscillatory shear simulations~\cite{yeh_glass_2019,bhaumik_role_2019}, showing that the transition from annealed to yielded behavior is found for parent temperatures in the vicinity of the dynamical crossover temperature $T_{\rm MCT}$.

\section{Conclusions}

All mesoscopic models of amorphous plasticity, starting from the early works by Argon and coworkers~\cite{bulatov_stochastic_1994a},  to the most recent space-resolved models~\cite{nicolas_deformation_2018}, postulate the existence of local thresholds, that are reshuffled during plastic events, and affected by local stress. These general assumptions have been guided by mainly two observations. First, plasticity results from the rearrangement of local packings, which suggests it is primarily controlled by local yielding thresholds. Second, it involves avalanche behavior, which implies long-range correlations mediated by elastic deformations. Mesoscopic models attempt to take these two observations into account by separating short scales (local packings, which determine local yield stresses) from large scales, i.e., the requirement of mechanical balance which couples elastically local stress levels. So, clearly, all mesoscopic models implement, out of necessity, the same general script. Yet, as soon as one looks at any level of detail beyond these rough guidelines, there is a flurry of such models and no consensus on the proper way to define a reliable one.

The problem is that, until now, as far as we know, no direct observation had been made of local yield thresholds in plastically deformed amorphous media. In the absence of such an observation, it has always remained a question whether the consideration of local yield thresholds was anything more that a simplifying assumption. And the key assumption about the effect of plasticity on local thresholds could only be build on guesswork: for example, it has never been clear whether it was meaningful to assume the existence of a well-defined postyield yield stress distribution. And no test was available for the prediction of models concerning the dynamics of their core state variable: the distribution of these local yield stresses.

The present work, although it does not, and cannot solve all these issues, brings hope that they can be addressed rigorously in the near future. It will not alleviate the need to rely on simplifying assumptions, and neither is it intended, to provide a recipe for the construction of mesoscopic model---a still distant goal. But it is the first to probe the rejuvenation process of atomic-scale yield stresses in a model glass undergoing deformation.  Thus, it brings direct observations that strongly constrain both model assumptions, especially about the effect of rejuvenation, and model predictions concerning the dynamics of local thresholds. Thanks to the focus on a system undergoing shear banding, i.e., comprising quite different local packings, we could demonstrate that there was indeed, a well-defined postyield yield stress distribution, quite insensitive to the initial ensemble, and we could access it numerically. This is evidently a key input for theories of amorphous plasticity.

Moreover, we found that rejuvenation was an unexpectedly rapid process, essentially requiring a single event to bring the local packing to a well-defined softness level, comparable to that of an IS obtained from a supercooled liquid around $T_{\rm MCT}$. This brings several important inputs to theories, specifically concerning our understanding of shear banding in amorphous systems. We showed indeed that the rejuvenation drop in yield stress causes the nucleation-like formation, immediately after the first yield events, of regions of well-defined softness. The rapidity of this process, resulting from the intensity and suddenness of rejuvenation softening, explains that, when starting from a well-relaxed glass, only a few plastic events suffice to locally erase the memory of the initial packing and achieve a soft state, which precipitates the shear banding instability. Also, the closeness of the postyield yield stress distribution to that of a liquid around the crossover region hint at the possibility to explain how the brittle-ductile transition depend on temperature, a very important practical issue.

By providing access to a relevant internal variable that characterizes the local mechanical properties, this work opens promising perspectives for describing nucleation and shear band dynamics in amorphous solids. It also establishes a stimulating link between mechanical and thermal properties, a necessary step to explain how the history of these out-of-equilibrium and disordered systems is encoded in their structure~\cite{Keim_memory_2018}.

These observations illustrate the remarkable ability of our tool to provide much needed insights on the local mechanisms of plasticity. We are convinced that a sustained exchange between such numerical probes, which provide both input and benchmarks, and the construction of mesoscopic models, will be instrumental in the coming years for the advancement of our fundamental understanding of amorphous plasticity. We only hope the present work is an early step along this promising route.

\begin{acknowledgments}
M.L. and S.P. acknowledge the support of French National Research Agency through the JCJC project PAMPAS under grant ANR-17-CE30-0019-01.
\end{acknowledgments}

\appendix

\section{Effective temperature scales \label{appendix_temperature}}

\subsection{Determination of the fictive Temperature}

\begin{figure}[h]
\includegraphics[width=0.95\columnwidth]{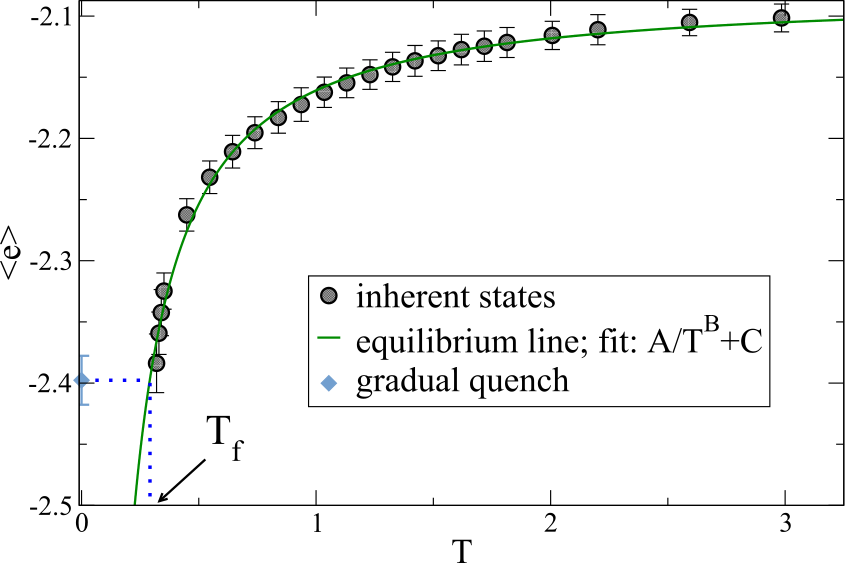}
  \caption{\label{fig:Tf} Inherent-state (IS) potential energy per atom as a function of temperature in equilibrated liquids (grey circles) and in our GQ system (blue diamond). The green line is a fit of the equilibrated liquid data. The fictive temperature $T_{\rm f}$ is estimated as that where the liquid IS energy extrapolates to that of the GQ system (see blue dotted lines).}
\end{figure}

The notion of fictive temperature $T_{\rm f}$ is classically used to map a nonequilibrium glass state on to an equilibrium liquid. Following Ref.~\cite{liu_correlating_2018}, the fictive temperature of our gradually quenched glass is defined using the inherent state potential energy: Namely, $T_{\rm f}$ is defined [see Fig. \ref{fig:Tf}] as the temperature at which the inherent potential energy of the glass equates that of equilibrated liquids. We obtain $T_{\rm f} = 0.291 \pm 0.007$.

\subsection{Determination of the Mode-Coupling Temperature}

\begin{figure}[h]
\includegraphics[width=0.95\columnwidth]{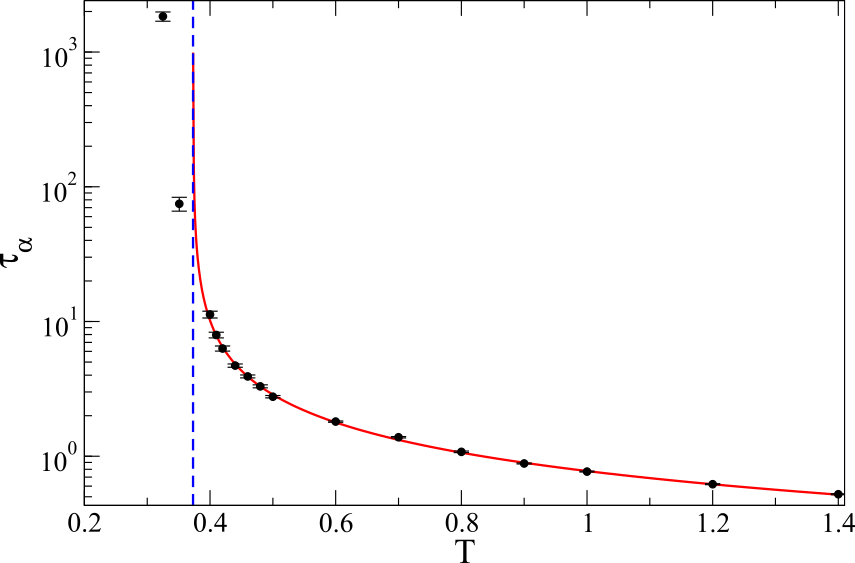}
  \caption{\label{fig:Tmct}Relaxation time $\tau_\alpha$ as a function of the equilibration temperature. The red line is a power-law fit of the data points for temperatures larger than $0.4$. The vertical blue line represents the location of estimated $T_{\rm MCT}$.}
\end{figure}

The Mode-Coupling Temperature $T_{\rm MCT}$ is determined from a functional fit of the relaxation time $\tau_{\alpha}$ in the dynamical regime for different temperatures $T$~\cite{schroder_crossover_2000,cavagna_supercooled_2009}. We
first compute the self-intermediate scattering function  $F_L(\textbf{q},t) = \langle \cos(\textbf{q} \cdot (\textbf{r}_j(t) - \textbf{r}_j(0))) \rangle$, where the subscript $L$ refers to the ``large'' particles, $\textbf{r}_j(t)$ is the position of the $j^{th}$ particle at time $t$ and $\langle \dots \rangle$ denotes the average over $j$ and the time origin. $F_L(\textbf{q},t)$ is averaged over $100$ samples, containing $1024$ atoms each, for every temperature. We define the relaxation time $\tau_\alpha$ as $F_L(\textbf{q}_{SF},t=\tau_\alpha) = \frac{1}{e}$ with $|\textbf{q}_{SF}| = 6.07$ corresponding to the primary peak of the static structure factor. As reported in Fig.~\ref{fig:Tmct}, $\tau_\alpha$, as computed for temperatures larger than $0.4$, is well fitted by the power-law relation $\tau_\alpha \propto (T-T_{\rm MCT})^\gamma$. We obtain $\gamma= -0.818 \pm 0.005$ and $T_{\rm MCT} = 0.373 \pm 0.001$.

\section{Strain measures \label{appendix_strain}}

\subsection{Hencky equivalent global shear-strain}

Our model glasses are deformed using Athermal Quasi-Static (AQS) simple shear loading up to linear strains $\gamma_{xy}=500\%$. The global strain is characterized within the finite-strain theory by using a Lagrangian description and choosing the Hencky's logarithmic strain definition~\cite{neff_axiomatic_2014}. The advantages of proceeding in such a way are threefold: It makes no use of small strain approximation, it allows us to eliminate rotations, and it will permit us to characterize macroscopic and microscopic strains using comparable quantities.

Given $\mathbf{F}$ as the deformation gradient, the logarithmic (Hencky) strain is $\mathbf{E}=\ln(\mathbf{U})$, where $\mathbf{U}$ is the right stretch tensor, i.e., the unique symmetric and positive-definite tensor such that $\mathbf{F}=\mathbf{R}\mathbf{U}$, with $\mathbf{R}$ a rotation. Since the right Cauchy-Green strain
\begin{equation}
\mathbf{F}^T\mathbf{F}=\mathbf{U}^2
\end{equation}
is positive definite, we may also write $\mathbf{E}=\frac{1}{2}\ln(\mathbf{U}^2)$. Moreover, $\mathbf{U}^2$ can be diagonalized, i.e., written as $\mathbf{U}^2=\mathbf{P}^{-1}\mathbf{D}\mathbf{P}$ with $\mathbf{D}$ diagonal and $\mathbf{P}$ the associated change of basis matrix. Finally, the invariants of $\mathbf{E}=\mathbf{P}^{-1} \ln(\mathbf{D}) \mathbf{P}$ are those of $\ln(\mathbf{D})$.

For two-dimensional simple shear deformation, the deformation gradient is:
\begin{gather}
\mathbf{F}=
  \begin{pmatrix}
   1 &
   \gamma \\
   0 &
   1
   \end{pmatrix}
\end{gather}
with $\gamma$ the linear strain, and the right Cauchy-Green strain,
\begin{gather}
\mathbf{U}^2=
  \begin{pmatrix}
   1 &
   \gamma \\
   \gamma &
   1+\gamma^2
   \end{pmatrix}
\end{gather}
It eigenvalues are $\lambda_\pm=\left(\frac{\gamma}{2}\pm\sqrt{1+\frac{\gamma^2}{4}}\right)^2$. Noting that $\lambda_+\lambda_- = 1$, since $\det(\mathbf{U}^2)=1$, the square-root of the second tensor invariant of $\mathbf{E}$, $E\equiv\frac{1}{4}|\ln(\lambda_+/\lambda_-)|=\ln\left(\frac{\gamma}{2}+\sqrt{1+\frac{\gamma^2}{4}}\right)$. This is the quantity we refer to in the manuscript as the macroscopic Hencky strain.

\subsection{Local strain computation}

The procedure for computing local strain tensor in our atomic system is the same as that previously employed in Refs.~\cite{patinet_connecting_2016,barbot_local_2018}. It relies on the evaluation of the local, coarse-grained, deformation gradient tensor $\mathbf{F}$.

In continuum mechanics $\mathbf{F}$ relates the position in the reference (undeformed) configuration to the current (deformed) configuration through the Cauchy-Born rule. For an atomic system under affine deformation this mapping relates the vectors $\mathbf{X}^{\alpha\beta}$ and $\mathbf{x}^{\alpha\beta}$ connecting atoms $\alpha$ and $\beta$ in the reference and current configurations (respectively) by:
\begin{equation}
\label{eq:Fdef}
{x}_{ij}^{\alpha\beta}=F_{ij}X_{ij}^{\alpha\beta}
\end{equation}
with $i$ and $j$ coordinate components. Since local strains are not affine in general (particularly in glasses~\cite{saw_nonaffine_2016}) the previous formula will no longer be exact for each individual atom-neighbor pair.

For two configurations separated by a small macroscopic strain interval, as proposed by Zimmerman \textit{et al.}~\cite{zimmerman_deformation_2009}, we define the atomic-level deformation tensor $F_{ij}^{\alpha}$ for atom $\alpha$ by minimizing the function
\begin{equation}
\label{eq:error}
B^{\alpha}=\sum_{\beta=1}^{n}\sum_{i=1}^{2}g(r^{\alpha\beta}_{0})(x_{i}^{\alpha\beta}-F_{ij}^{\alpha}X_{j}^{\alpha\beta})^{2},
\end{equation}
where the sum runs over the $n$ nearest neighbors of $\alpha$, and where $g(r^{\alpha\beta}_{0})$ is a smooth weighting function that only depends on the distance $r^{\alpha\beta}_{0}$ between atoms $\alpha$ and $\beta$ in the reference configuration~\cite{hinkle_coarse_2017}. Thus $F_{ij}^{\alpha}$ is the best fit of Eq.~(\ref{eq:Fdef}), in the least-squares sense, for an atom in its cage. We choose for $g$ an octic polynomial function~\cite{lemaitre_structural_2014}. This function has a single maximum and continuously vanishes at a coarse graining length $R_{CG}=5$:
\begin{equation}
\label{eq:smooth_weighting_function}
g(r) =
\begin{cases}
\frac{15}{8\pi R_{CG}^2}(1-2(\frac{r}{R_{CG}})^4+(\frac{r}{R_{CG}})^8), &\text{for } r<R_{CG}\\
0, &\text{otherwise}.
\end{cases}
\end{equation}
The minimization of Eq.~\ref{eq:error} with respect to $\mathbf{F}^{\alpha}$ gives
\begin{equation}
\label{eq:sol}
F_{ij}^{\alpha}=\sum_{k=1}^{2}Y_{ik}Z_{jk}^{-1},
\end{equation}
with
\begin{equation}
Y_{ik}=\sum_{\beta=1}^{n}(x^{\alpha\beta}_{i}X^{\alpha\beta}_{k})g(r^{\alpha\beta}_{0})  \textrm{ and } Z_{ik}=\sum_{\beta=1}^{n}(X^{\alpha\beta}_{i}X^{\alpha\beta}_{k})g(r^{\alpha\beta}_{0}).
\end{equation}

This procedure is used to compute the strain gradient for each atom $\mathbf{F}$ between macroscopic configuration separated by small strain intervals $\Delta\gamma=0.01$. For larger $\Delta\gamma$, $\mathbf{F}$ is numerically integrated by multiplying the deformation gradient tensors $\mathbf{F}=\mathbf{F}_{n}\mathbf{F}_{n-1}\cdot\cdot\cdot\mathbf{F}_{2}\mathbf{F}_{1}$ where $\mathbf{F}_{i}$ is the deformation gradient tensors between states $i$ and $i+1$ separated by $\Delta\gamma=0.01$.

The local strain is reduced to a scalar following the same procedure as for the global strain $E$, which involves the numerical diagonalization of each $\mathbf{F}^T\mathbf{F}$. Finally, the atomic-level strain $\epsilon$ is defined as the second tensor invariant of $\frac{1}{2}\ln(\mathbf{F}^T\mathbf{F})$.

The $\epsilon$ field is then evaluated on a square grid $N^2$ by assigning to each grid point the value computed for the closest corresponding atom. The regular grid lattice constant equals to $R_{sampling}=L/39\approx~R_{cut}$ where $L$ is the dimension of the initially square simulation box and $R_{cut}=2.5\sigma$ is the cutoff interatomic potential. The grid deforms affinely with the overall applied strain in simple shear. The local yield stresses are computed on the same grid.

\section{Local properties \label{appendix_localprop}}

\begin{figure}
\begin{center}
\includegraphics[width=0.95\columnwidth]{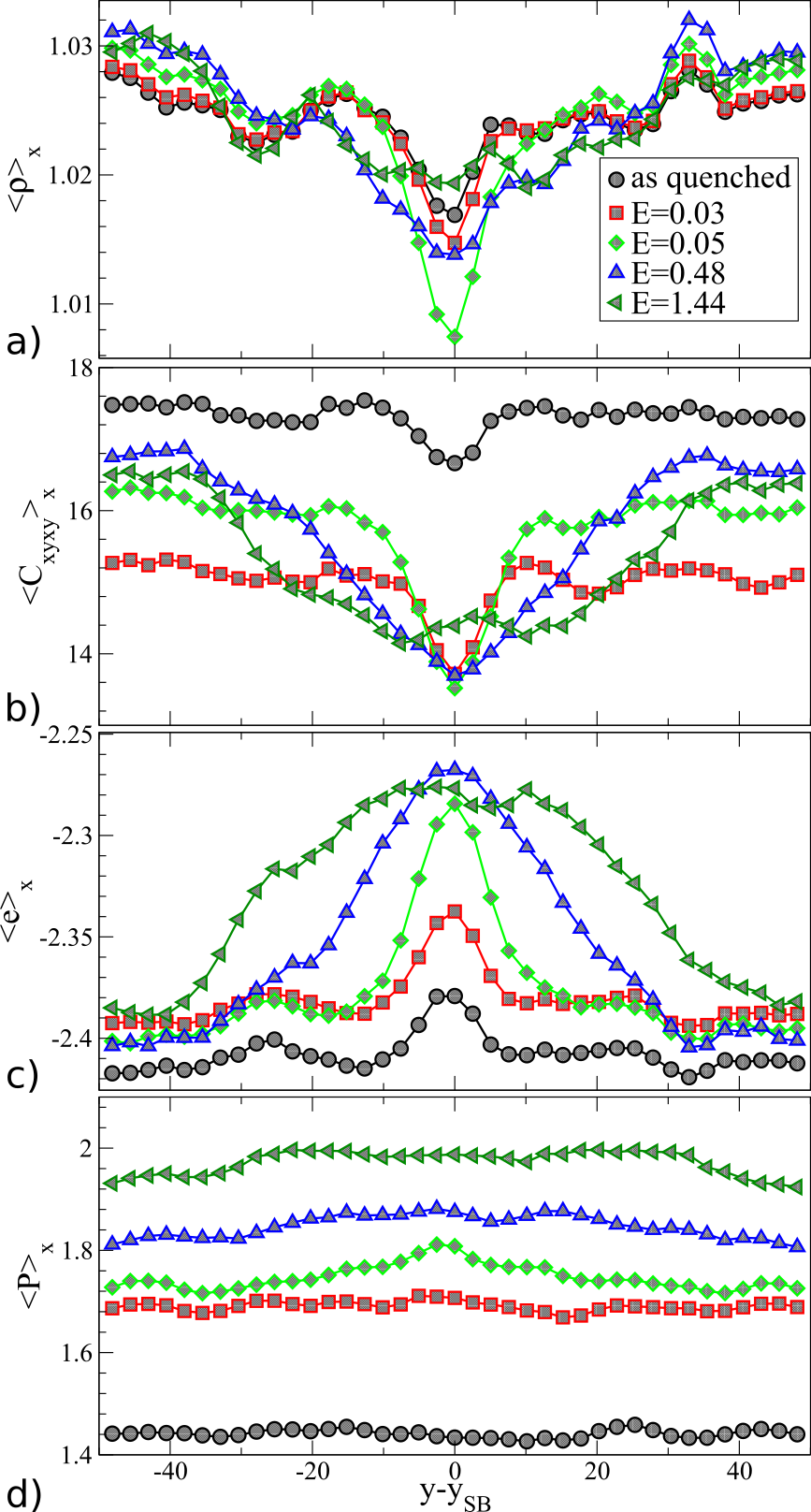}
\caption{\label{fig:profiles} Local averages computed along the shear direction $x$ for different strain levels $E$ as a function of the vertical position $y$ with respect to the shear band position $y_{SB}$: a) density $\langle\rho\rangle_{x}$, b) shear modulus in the loading direction $\langle C_{xyxy}\rangle_{x}$, c) potential energy per atom $\langle e\rangle_{x}$, and d) pressure $\langle P\rangle_{x}$.}
\end{center}
\end{figure}

\subsection{Profiles through the shear band}

In Fig.~\ref{fig:profiles}, we report the average profiles of density $\rho$, elastic modulus $C_{xyxy}$, potential energy per atom $e$ and pressure $P$ for different macroscopic strain levels. As described in the manuscript, we restrict this analysis to samples featuring only horizontal shear bands. The profiles are centered on the shear band ordinate $y_{SB}$, which is defined at that where the $x$-averaged local strain is maximum for $E=0.1$.

The values of $\langle\rho\rangle_{x}$, $\langle C_{xyxy}\rangle_{x}$, and $\langle e\rangle_{x}$ in the shear band and the outer medium are visibly contrasted: The accumulation of plastic rearrangements brings the system to a less dense, less rigid, and less stable state. Note, however, that all profiles eventually become homogeneous in the stationary flow state.

Because of the constant volume loading protocol, the average of $\rho$ is by definition equal to the mean system density. The local density in the band thus reaches transiently a minimum around $E\simeq 0.05$ before increasing again towards the mean density, its eventual value at large macroscopic strains.

The evolution of the elastic moduli is somewhat curious. While it softens in the band, $\langle C_{xyxy}\rangle_{x}$ presents an overall decrease during the early stages of loading up to the peak stress. This is expected since the approach to plastic instabilities reduces the macroscopic elastic modulus~\cite{maloney_universal_2004}. The moduli in the outer, elastic region, then rapidly rises as the stress decreases from its peak, presumably because the density of near threshold zone diminishes.

The presence and broadening of the shear band can also be detected by examining the increase in $\langle e\rangle_{x}$. This increase has already been observed in a very similar atomic system \cite{shi_evaluation_2007}, where it has been interpreted as a rise in effective temperature through Shear-Transformation-Zone theory.

It is interesting to note that the shear band position at $E=0.1$, which we take as reference, correlates with various local properties of the quenched state. Note, finally, that if the global pressure increases with plastic deformation (see Fig.~1 in the manuscript), it is almost homogeneous along these average profiles due to mechanical equilibrium as reported in Fig.~\ref{fig:profiles}d.

\subsection{Variation of local yield stresses with pressure}

\begin{figure}
\begin{center}
\includegraphics[width=0.95\columnwidth]{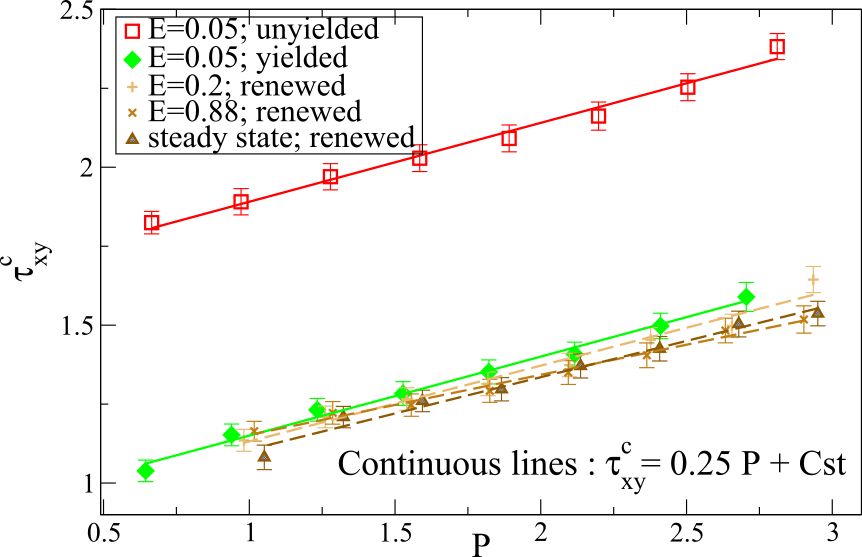}
\caption{\label{fig:tauc_vs_p} Local yield stress $\tau^{c}_{xy}$ as conditioned by local pressure $P$ in yielded (green) and unyielded (red) regions at $E=0.05$ and for renewed sites (shades of brown) at different macroscopic strains.}
\end{center}
\end{figure}

We report in Fig.~\ref{fig:tauc_vs_p} the average of $\langle\tau_{xy}^{c}\rangle$ as a function of the local pressure $P$ in the yielded and nonyielded regions at $E=0.05$ and for renewed states (see manuscript) for different macroscopic strains. This plot clearly shows that, at the same $P$, unyielded and yielded regions present different yield stresses. Moreover, in either ensemble, the local yield stress increases nearly linearly with local pressure. This result is in agreement with the variation of the yield surface obtained numerically for a metallic glass in Ref.~\cite{lund_yield_2003}, where pressure or normal stress give results intrinsically similar to a Mohr-Coulomb criterion. The slopes of the $\langle\tau_{xy}^{c}\rangle$ relations differ by merely $10\%$. We therefore choose to adjust the slope on all the data. We find an empirical relationship for yielded and unyielded sites: $\langle\tau_{xy}^{c}\rangle=0.25~P + A$ where $A$ is a constant.

Using the identified slope of the local relation between $\tau_{xy}^{c}$ vs $P$, we can now take into account the effect of local pressure and correct the data in Fig. 4b of the manuscript. We calculate the threshold distribution of unyielded sites at $E=0.05$ from the corresponding thresholds in the quenched state as: $\tau_{xy}^{c}=\tau_{xy}^{c}(E=0 | E=0.05, \epsilon<\langle\epsilon^{*}\rangle/8) + 0.25 \Delta P $ where $\Delta P$ is the variation of local pressure on given sites between $E=0$ and $E=0.05$. The corrected curves (green triangles in Fig.~4b in the manuscript) quantitatively reproduce the distributions $\tau_{xy}^{c}$ of unyielded sites at $E =0.05$. This establishes that changes in the local yield stress for small deformations, mainly outside of the band, can be attributed both to the depletion of weak sites and, in the case of constant volume simulations, to the increase in local pressure.

The relation between $\langle\tau_{xy}^{c}\rangle$ and $P$ for the renewed states at different macroscopic strains still shows a nearly linear dependency. It, however, slightly deviates from the relation found for the yielded sites at $E=0.05$.

%\bibliography{biblio_shearband_BLLVP}

%merlin.mbs 2010-03-15 4.21a (PWD, AO, DPC)
%Control: key (0)
%Control: author (8) initials jnrlst
%Control: editor formatted (1) identically to author
%Control: production of article title (-1) disabled
%Control: page (0) single
%Control: year (1) truncated
%Control: production of eprint (0) enabled
%

\end{document}